\documentclass[amssym,amsmath,amsfonts,aps]{appolb}

\usepackage{graphicx,mathrsfs,flafter}
\usepackage{color}
\usepackage{hyperref}

\begin{document}
\title{Phase transitions at finite 
density\thanks{Presented at the HIC for FAIR Workshop and XXVIII Max Born Symposium \emph{Three days on Quarkyonic Island}, Wroc\l aw, 19-21 May, 2011}%
}

\author{B. Friman
\address{%
GSI Helmholtzzentrum f\"ur Schwerionenforschung\\ D-64291
Darmstadt, Germany}
}

\date{\today}
\maketitle

\begin{abstract} 

I discuss the analytic structure of thermodynamic quantities for complex values of thermodynamic variables within Landau theory.  In particular, the singularities connected with phase transitions of second order, first order and cross over types are examined. A conformal mapping is introduced, which may be used to explore the thermodynamics of strongly interacting matter at finite values of the baryon chemical potential $\mu$ starting from lattice QCD results at $\mu^{2}\leq 0$.  This method allows us to improve the convergence of a Taylor expansion about $\mu=0$ and to enhance the sensitivity to physical singularities in the complex $\mu$ plane. The technique is illustrated by an application to a second-order transition in a chiral effective model.
\end{abstract}

\section{Introduction}
The properties of strongly interacting matter 
at finite temperature, $T\neq0$, and baryon density, $n_{B}\neq 0$, have been studied extensively in recent years. The phase diagram has been 
explored experimentally in heavy-ion collisions and examined theoretically in 
models as well as in
first-principle calculations of Quantum Chromodynamics (QCD).
In the non-perturbative regime of QCD, Lattice Gauge Theory (LGT) provides a powerful tool for computing the thermodynamic properties of strongly interacting matter. 
However, the so called fermion sign problem prohibits a straightforward application of LGT at nonzero (real) values of the baryon chemical potential. 

Various approaches have been
developed, to sidestep this problem~(for a review see Ref.~\cite{Philipsen:2005mj}). These generally involve an extrapolation from ensembles, where the fermion determinant is positive definite, e.g. from $\mu=0$ or from imaginary $\mu$. Using such techniques, thermodynamic functions can be computed for small real $\mu/T$, i.e.,  for small baryon densities. 

The extrapolation from the $\mu=0$ ensemble can be performed, 
e.g. by means of a Taylor expansion~\cite{Allton:2002zi} in $\mu/T$.
Without refinements, this method is applicable only within the radius
of convergence of the series, $R_{\mu}$. For imaginary $\mu$, the fermion determinant is strictly 
positive definite and consequently systematic LGT calculations are possible~\cite{Roberge:1986mm,Alford:1998sd,deForcrand:2002ci,D'Elia:2002gd}. 
An analytic continuation to real $\mu$ by means of a polynomial
is valid only within the convergence radius of the Taylor
expansion~\cite{deForcrand:2002ci}. Thus, both methods are
restricted in their applicability by $R_{\mu}$. 

The radius of convergence of a Taylor series is limited by the distance to the closest singularity. 
Conversely, the convergence properties of a power series provides information 
on the singularities of the original function. Of particular interest are the singularities that are related
to a phase transition~\cite{Hemmer52,Stephanov:2006dn,Karbstein:2006er}. The analytic structure in the complex $\mu$ plane in finite systems was explored in a simple exactly solvable model in Ref.~\cite{Stephanov:2006dn}. 
In this talk, I focus on systems in the thermodynamic limit, and do not discuss finite-size effects.

The convergence properties of a Taylor series in $\mu$ 
has been studied in model calculations~\cite{Karbstein:2006er,Wagner:2010}. 
It was found that on the order of 20 terms or more are needed
to obtain reliable information on the structure of the phase diagram
for such models. This is well beyond what is presently available in LGT
calculations~\cite{Gavai:2008zr,Schmidt:2010xm}. Here more powerful methods
for the analysis of a truncated power series, based e.g. on the
Pad\'{e} conjecture (for an application to QCD in Ref.~\cite{Lombardo:2005ks}) or on analytic continuation using conformal mappings~\cite{Skokov:2010uc}, may prove useful.

In this talk I first discuss the singularities of the order parameter for different types of phase transitions within the Landau theory.  Subsequently, I briefly describe the analytic structure in the complex $\mu$ plane in an effective model, the quark-meson model~\cite{qm}. Finally, I present a method using a conformal mapping for determining the approximate location of a second-order critical point, given a finite number of terms in a series expansion of a thermodynamic function. 

\section{Analytical structure of thermodynamic functions}

We now discuss the analytical structure of thermodynamic functions in systems with a phase transition.  The general features are illustrated in a transparent framework, the Landau theory of phase transitions. Consider first a second order phase transition, described by the Hamiltonian density\footnote{Since we consider only uniform consdensates, gradient terms do not contribute.}
\begin{equation}
\mathcal H\left[\sigma\right] = \frac{a}{2}\sigma^{2}+\frac b4\sigma^{4}+\frac c6\sigma^{6} -h\sigma,
\end{equation}
where $\sigma$ is the order parameter of the phase transition and $h$ is a symmetry-breaking external field.
The parameter $a$ depends on temperature and chemical potential, and vanishes at the second order critical point\footnote{The critical point  $a_{\rm c}=0$ may correspond to a critical line in the $T-\mu$ plane. In this discussion $a$ is a proxy for the complex chemical potential.} ($a_{\rm c}=0$), while $b$ is a positive constant and $c$ is introduced for later use. For now we put $c=0$.

We consider the analytic structure of the order parameter as a function of a complex $a$.  The order parameter as a function of $a$ is obtained by solving the gap equation
\begin{equation}\label{gap}
\frac{\partial \mathcal H }{\partial \sigma}=a \sigma + b \sigma^{3}-h=0.
\end{equation}

Consider first a system with exact symmetry, i.e. $h=0$. For real $a$ below the critical point (Re$\left[a\right]<0$, Im$\left[a\right]=0$), the symmetry is spontaneously broken and the order parameter is non-zero. At the critical point, $a_{\rm c}=0$, the three solutions of the gap equation coincide; this is a branch point of a cut along the positive real $a$-axis for the two solutions with non-zero order parameter\footnote{In general a singularity is located at a point where at least two solutions come together. The solutions of the gap equation form Riemann surfaces that are joined at the cuts.}. 
A Taylor expansion about a point in the broken symmetry phase, $a_{0}$ with Re$\left[a_{0}\right]<0$ and Im$\left[a_{0}\right]=0$, converges within the radius equal to the distance to the critical point, $R_{a}=| a_{0}-a_{\rm c}|$.  

We now turn to the case with an explicit symmetry breaking term, i.e. $h\neq 0$, while $b$ is still positive. In this case there is no true phase transition; the order parameter is always non-zero. The second order transition is replaced by a cross over transition. The singular points in the complex $a$ plane are  obtained by solving a system of equations consisting of the gap equation and its derivative $\partial^{2}\mathcal H/\partial \sigma^{2}=0$ for $a$. 
\begin{figure}[t]
\centerline{\includegraphics*[width=4.5cm]{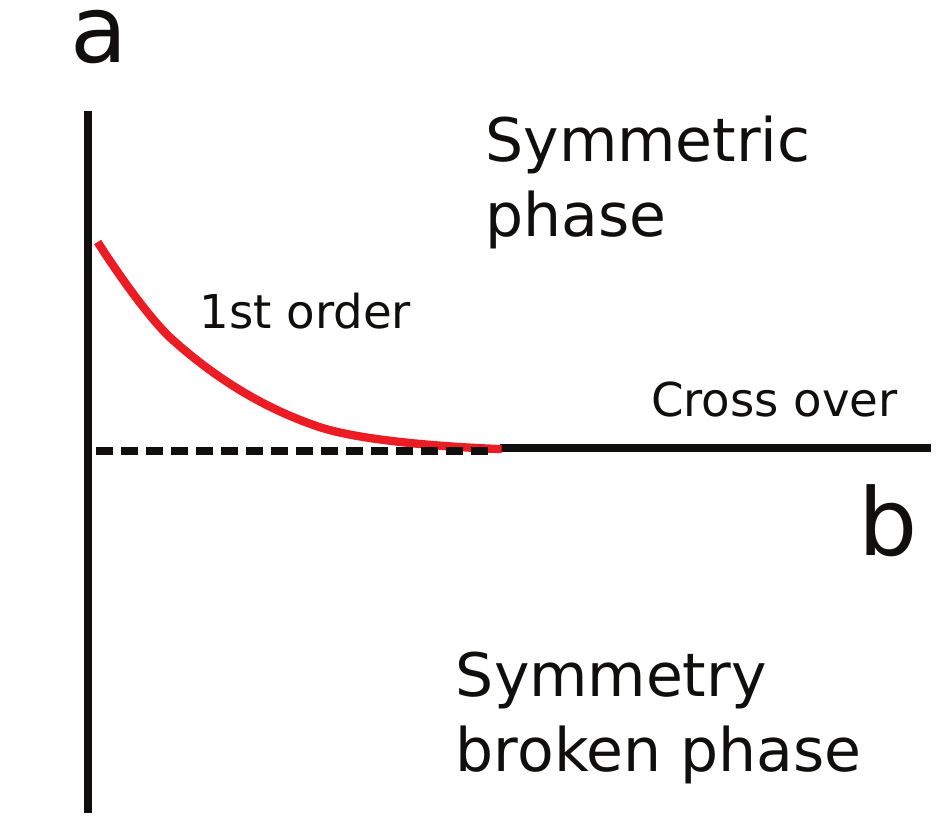}}
\caption{\label{landau-pd}Schematic phase diagram of Landau theory, with $h\neq 0$.}
\end{figure}

There are three solutions $a_{n}=-r\, e^{-2n i \pi/3}$, where $r=3(bh^{2}/4)^{1/3}$ and $n=0,1,2$; the two complex ones are relevant for the cross-over transition. The singularity corresponding to the critical point of the second order transition is, for $h\neq 0$, split into two singularities at complex conjugate values of $a$, $a_{\rm co}=a_{1}$ and $a_{\rm co}^{*}=a_{2}$. Both singularities are branch points of the physical solution, which minimizes $\mathcal H$ for real $a$, and thus limits the radius of convergence of a Taylor expansion about $a_{0}$. The singularity at $a=-r$ is closer to $a_{0}$, but is located on an unphysical Riemann surface, and therefore does not affect the convergence properties.
 
To describe a first-order transition, we allow also $b$ to vary, and take $c$ to be a positive constant to insure stability.  For positive $b$, the transition is second order or of the cross-over type, depending on whether the symmetry-breaking field $h$ vanishes or not. However, for negative $b$, the gap equation
\begin{equation}\label{gap5}
\frac{\partial \mathcal H }{\partial \sigma}=a \sigma + b \sigma^{3}+ c \sigma^{5}-h=0.
\end{equation}
has, for small  positive values of $a$, five real solutions, corresponding to three minima and two maxima of $\mathcal H$. Consequently, there is a first-order phase transition, with a discontinuous change of the order parameter. For $h\neq 0$, the corresponding phase diagram in the $a-b$ plane is shown schematically in Fig.~\ref{landau-pd}. The cross-over transition, defined by the maximum of the order-parameter susceptibility, is shifted to a positive $a\sim h^{4/5}$. Similarly, the critical endpoint, where the first order transition ends, is shifted from the tricritical point $(h=0)$ located at $a=b=0$ to $a_{\rm cep}\simeq 2.28 \,h^{4/5},b_{\rm cep}\simeq-2.25\, h^{2/5}$. 
\begin{figure}[b]
\centerline{\includegraphics*[width=6cm]{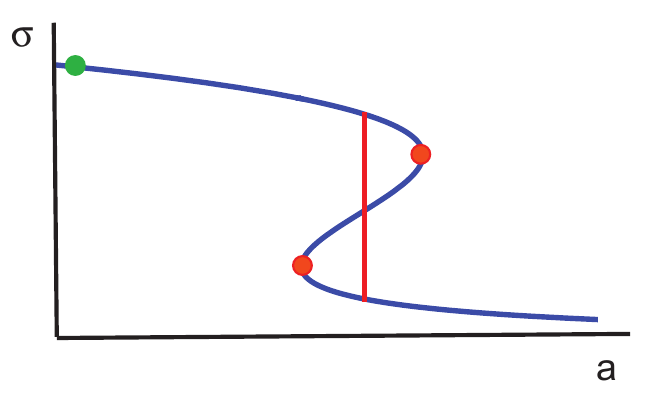}}
\caption{\label{maxwell}Schematic plot of the order parameter as a function of $a$ for a first-order transition. The Maxwell construction is indicated by the vertical line.}
\end{figure}
\begin{figure}[t]
\centerline{\includegraphics*[width=6cm]{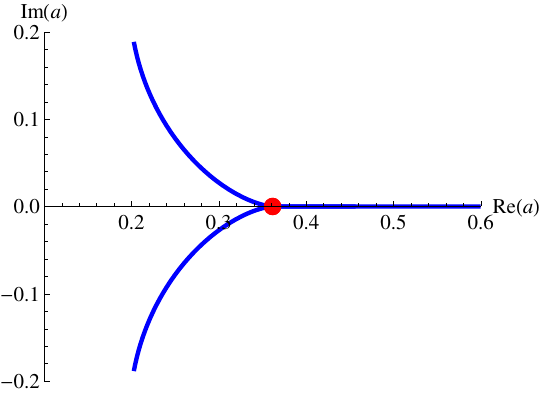}}
\caption{\label{sing-a-plane}The movement of the relevant singularities as functions of $b$.}
\end{figure}

The dependence of the order parameter as a function of $a$ for a first-order transition is shown schematically in Fig.~\ref{maxwell}. The location of the phase transition is indicated by the Maxwell construction, the vertical line in the plot. Two solutions of the gap equation coincide at the singular points, where $\partial\sigma/\partial a=\infty$, marked by dots. These are the spinodal points, where the corresponding solution becomes unstable. Between the spinodal points and the Maxwell construction, the solutions are metastable. 

The convergence radius of an expansion about e.g. the point close to the $y$-axis in Fig.~\ref{maxwell} is determined by the spinodal point of the same solution. Thus, neither the Maxwell construction nor the other spinodal point affect the convergence properties of the Taylor expansion, although they are closer to the expansion point\footnote{We note that the analytic structure near a first-order transition may in fact be more complex than implied by the mean-field arguments employed here~\cite{w-klein}. Moreover, in finite systems, the closest singularities are the Yang-Lee zeroes~\cite{Yang:1952be}, which, as discussed in~\cite{Stephanov:2006dn}, are reflected in the convergence properties of a Taylor expansion at large orders.}.  In Fig.~\ref{sing-a-plane} we show the movement of the singularities in the complex $a$ plane as a function of $b$. For $b<0$, the transition is first order and the relevant singularity is located on the real axis. At the critical end point, the first-order transition ends, and is replaced by a cross-over transition. There the singularity splits into two, which move into the complex $a$ plane.

\begin{figure}[t]
\centerline{\includegraphics*[width=7cm]{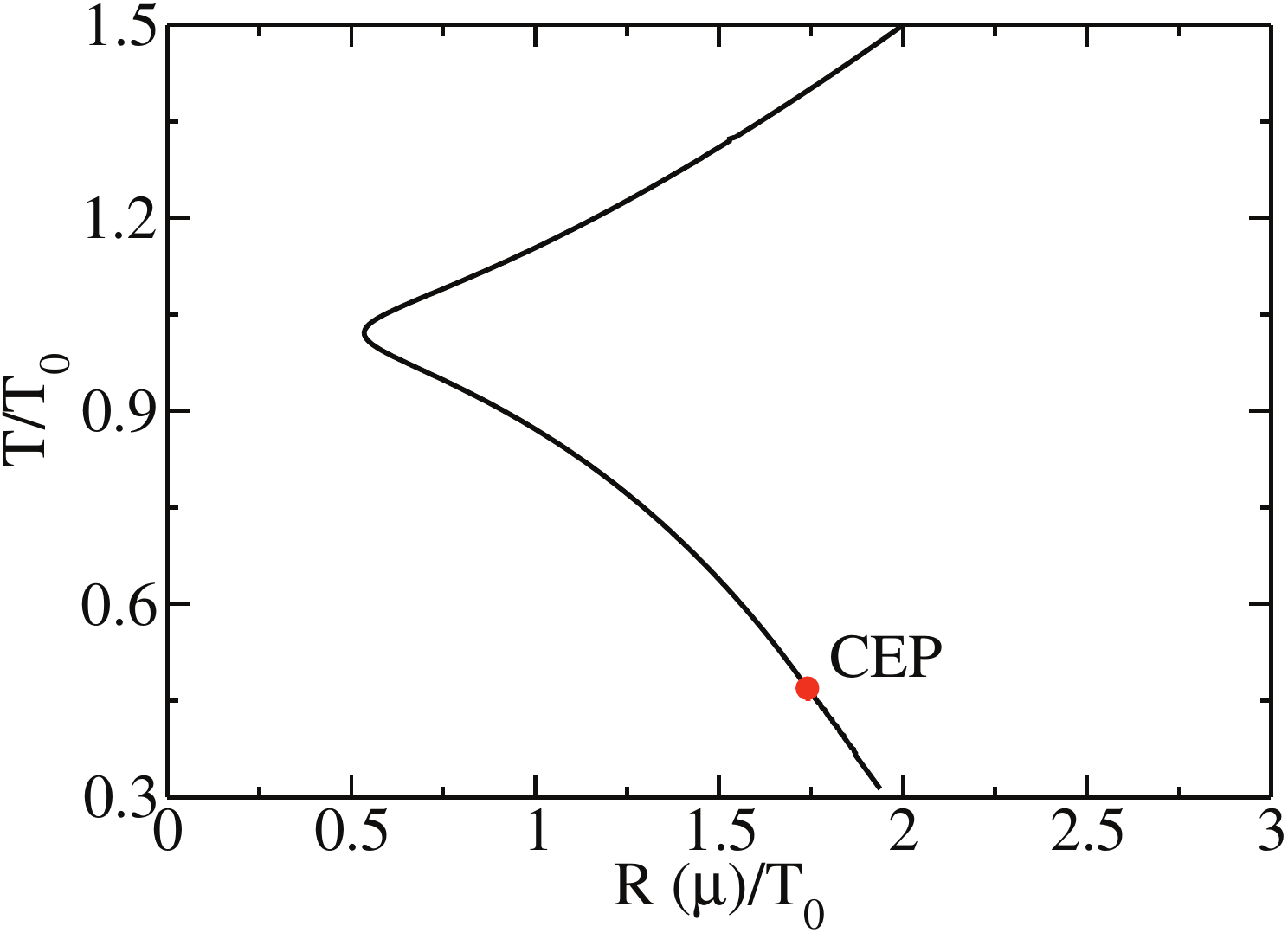}}
\caption{\label{tmuplane_norm}The radius of convergence in $\mu$ related to the first-order and cross-over transitions in the QM model for a non-zero quark mass.}
\end{figure}

The discussion of the analytic structure in the complex $a$ plane and the corresponding convergence properties can be directly adapted to an expansion in 
the complex $\mu$ plane. Thus, the analytic structure connected with a phase transition is reflected in the convergence properties of a Taylor  expansion in $\mu$. This can be utilized to locate the critical point approximately given a finite number of terms in the expansion. In Fig.~\ref{tmuplane_norm} we show the radius of convergence obtained from the analytic structure in the complex $\mu$ plane for the quark-meson model for a non-zero quark mass. In this case the radius of convergence is minimal close to the pseudo-critical temperature at $\mu=0$ and is due to the cross-over transition.

\section{Conformal mapping}

\begin{figure}[b]
 \centerline{ \includegraphics*[width=7cm]{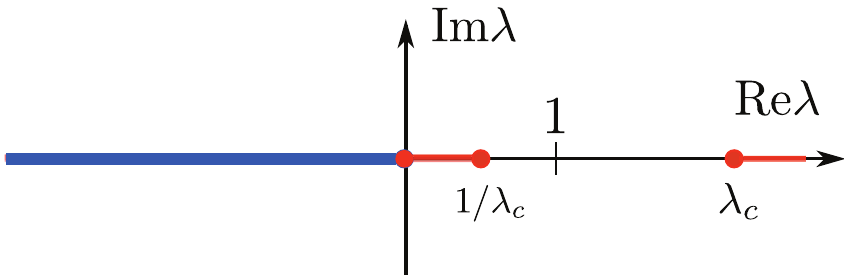}}
  \caption {\label{plane_fug3}
The analytic structure of the quark-meson model in the chiral limit at a temperature above the tricritical point, where there is a second-order transition at a finite density.
The branch point at $\lambda_{\rm c}=e^{\mu_{\rm c}/T}$ is the critical point of the second-order transition.  
     }
\end{figure}

\begin{figure}[t]
 \centerline{ \includegraphics*[width=3.5cm]{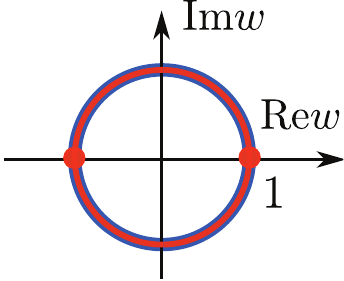}\quad\includegraphics*[width=3.5cm]{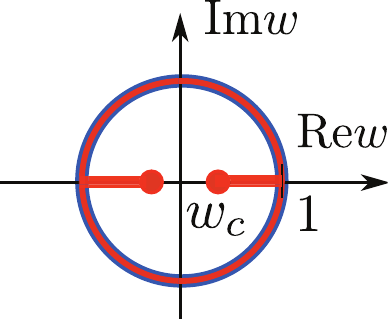}\quad\includegraphics*[width=3.5cm]{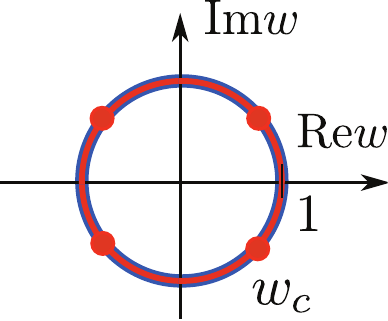}}
  \caption {The unit circle in the $w$ plane, the result of the conformal mapping~(\ref{conf-map}) of the cut $\lambda$ plane in Fig.~\ref{plane_fug3}. The middle and right panels show the analytic structure obtained with the mapping (\ref{lambda-g}) for $\lambda_{g}>\lambda_{c}$ and $\lambda_{g}<\lambda_{c}$, respectively.
     }
  \label{unit-w-circle}
\end{figure}

The analytic structure of the quark-meson model in the complex fugacity $\lambda=e^{\mu/T}$ plane is show in Fig.~\ref{plane_fug3} for temperatures above the critical end point and in the chiral limit.
There are two cuts associated with the critical point $\mu_{\rm c}$ and at its mirror image at $-\mu_{\rm c}$ that start at $\lambda_{\rm c}$ and $1/\lambda_{\rm c}$, respectively. Furthermore, there is a cut on the negative real $\lambda$ axis, which is due to the singularities of the Fermi function. The cut $\lambda$ plane is mapped on to a unit circle by the conformal mapping
\begin{equation}
w(\lambda)=\frac{\sqrt{\lambda \lambda_{\rm c}-1}-\sqrt{\lambda_{\rm c}-\lambda}}{\sqrt{\lambda \lambda_{\rm c}-1}+\sqrt{\lambda_{\rm c}-\lambda}}\, .
\label{conf-map}
\end{equation}
The branch points at $\lambda_{\rm c}$ and $1/\lambda_{\rm c}$ are mapped onto $w=1$ and $w=-1$ respectively, while the corresponding cuts are
mapped onto the circumference of the unit circle, as shown in the left panel of Fig.~\ref{unit-w-circle}. 
A Taylor expansion in $w$ converges everywhere within the circle. Hence, by applying the inverse mapping to this Taylor expansion, one obtains an expansion which converges in the complete cut $\lambda$ or equivalently $\mu$ plane. We note that $n$ terms of the original expansion are sufficient to compute the $n$ first Taylor coefficients of the $w$ expansion.

In applications, e.g. to lattice QCD, the location of the critical point in the $\mu$ plane is not known. Hence, a strategy by which one can find the (approximate) location of singularities, given a finite number of terms in the Taylor expansion, is needed. 
Below we present a method which is useful for locating a second-order phase transition, i.e., a critical point on the real $\mu$ axis. In the case of QCD, this could be a critical endpoint, where the crossover transition ends and is replaced by a first-order transition. In order to deal with a cross-over transition, where the singularity is located at complex $\mu$, a generalization of this method would be required.
\begin{figure}[t]
 \centerline{\includegraphics*[width=5.7cm]{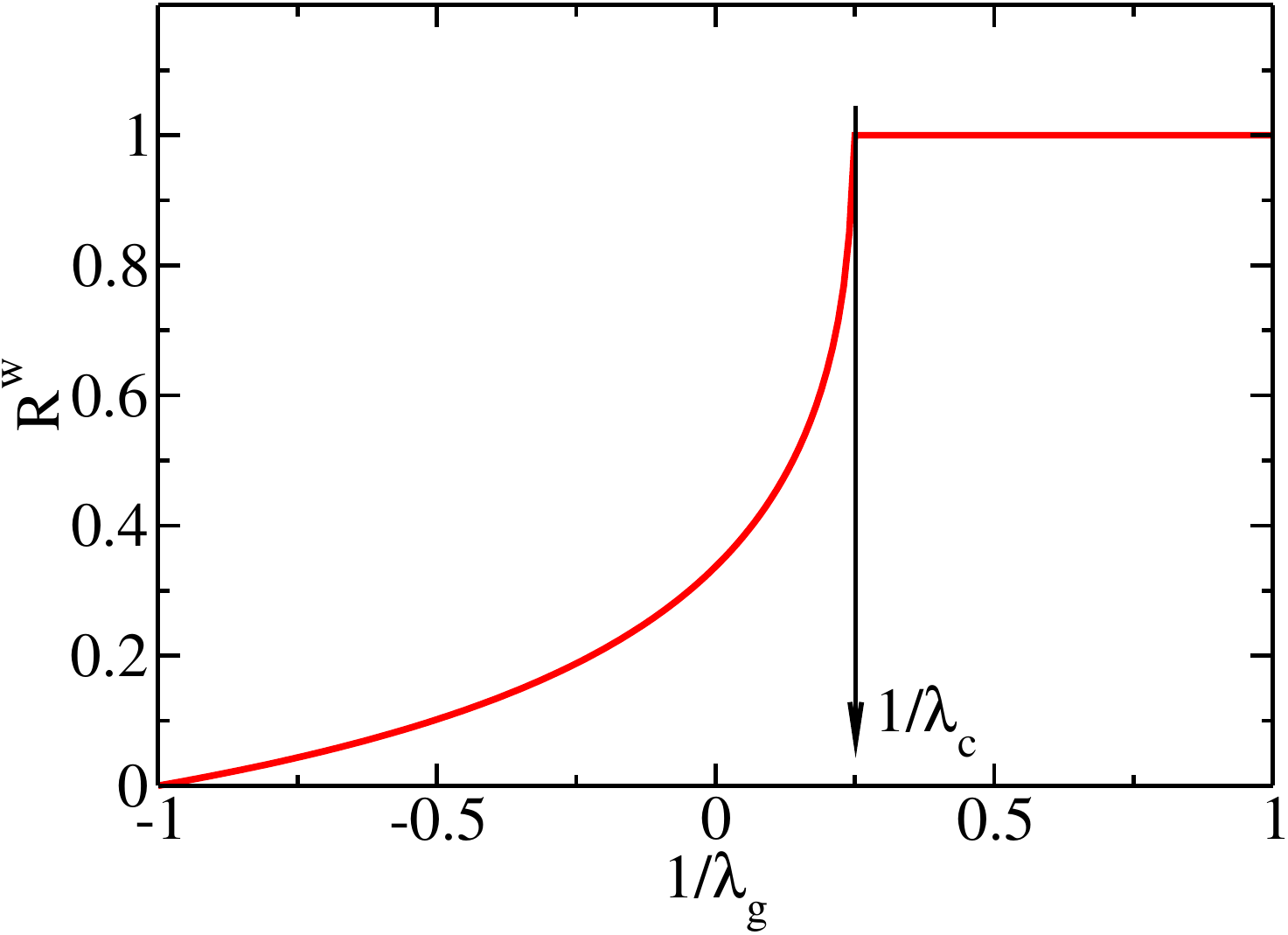}\quad \includegraphics*[width=6.cm]{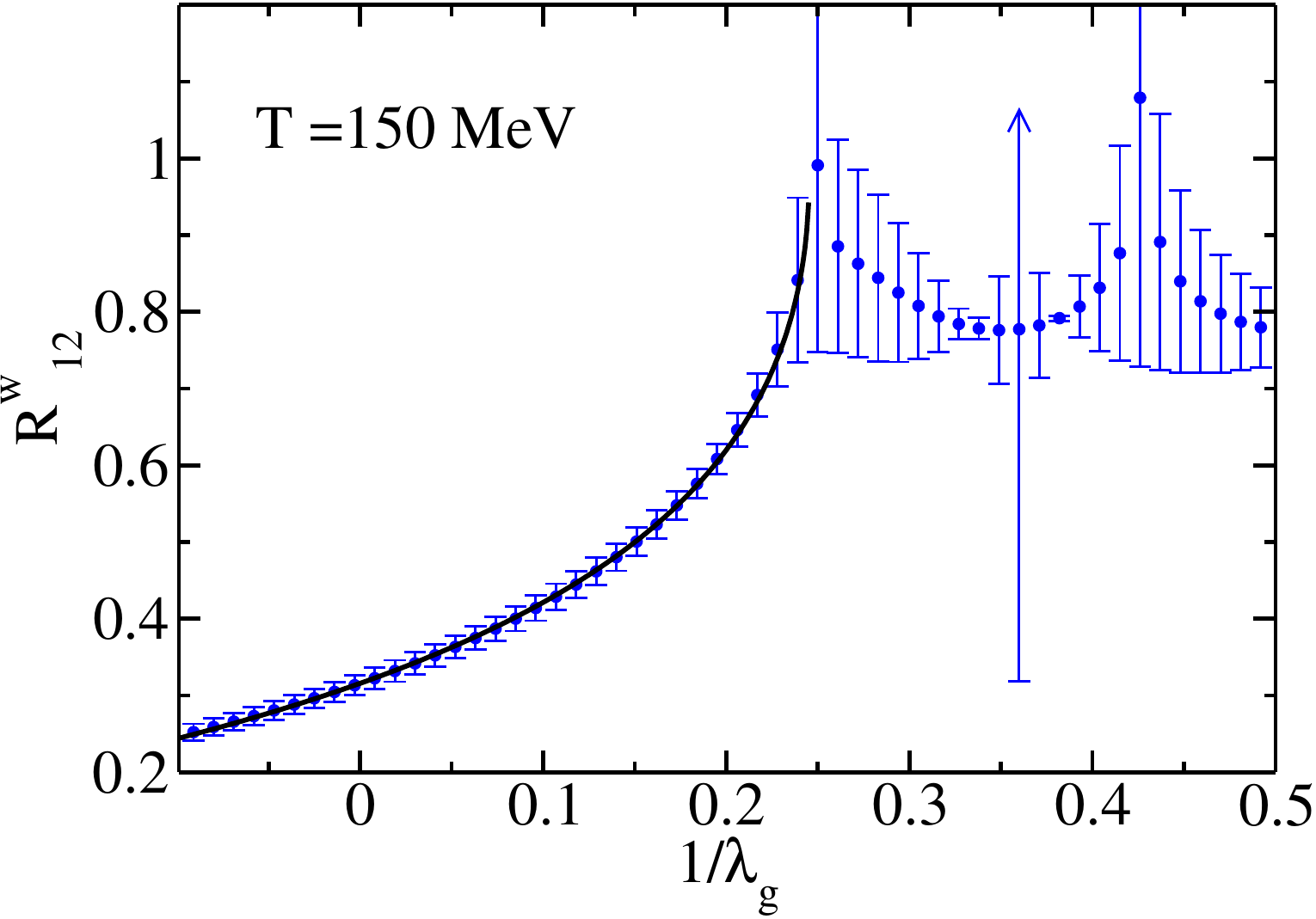}}
  \caption {The radius of convergence $R_n^w$ in the $w$ plane as a function of $1/\lambda_{\rm g}$. The left panel shows $w_{g}(\lambda_{c};\lambda_{g})$ for $\lambda_{c}=4$, while the right panel shows the $R^{w}_{n}$ computed in the QM model for $n=12$ (dots). The error bars show the deviation of the radius of convergence $R_{12}^w$ from that obtained with $n=10$.   
The solid line show the result of fitting the {\em Ansatz} $R^{w}(\lambda_{\rm g})$.
     }
  \label{fig:Rconv}
\end{figure}

First we define a map of the type (\ref{conf-map}), replacing the critical fugacity $\lambda_{\rm c}$ by a parameter  $\lambda_{\rm g}$
\begin{equation}\label{lambda-g}
w_{\rm g}(\lambda;\lambda_{\rm g})=\frac{\sqrt{\lambda \lambda_{\rm g}-1}-\sqrt{\lambda_{\rm g}-\lambda}}{\sqrt{\lambda \lambda_{\rm g}-1}+\sqrt{\lambda_{\rm g}-\lambda}}\, .
\end{equation}
The analytical structure of the thermodynamic function in the $w_{\rm g}$ plane now depends on the value of $\lambda_{\rm g}$ relative to $\lambda_{\rm c}$. For $\lambda_{\rm g}>\lambda_{\rm c}$, the branch points associated with the critical point, 
$\lambda_{\rm c}$ and $1/\lambda_{\rm c}$, are mapped onto points inside the unit circle at $w=\pm w_{\rm c}=\pm w_{\rm g}(\lambda_{\rm c};\lambda_{\rm g})$ 
(see Fig.~\ref{unit-w-circle}, middle panel).
Since $w=\pm w_{\rm c}$ are the singularities that are closest to the origin, the radius of convergence of the Taylor expansion in $w$ is given by $R_{w}=w_{\rm c}$.
If, on the other hand,
$\lambda_{\rm g}<\lambda_{\rm c}$, the critical point is mapped onto the
circumference of the unit circle (Fig.~\ref{unit-w-circle}, right panel) and consequently the radius of convergence equals unity, $R_{w}=1$. The expected dependence of $R_{w}$ on $\lambda_{g}$ is shown in the left panel of Fig.~\ref{fig:Rconv}. In the ideal case, where $R_{w}$ is known exactly, any point on the curve with $R_{w}<1$ could be used to determine $\lambda_{c}$ using the inverse mapping to $w=R_{w}$.
When only a finite number of terms of the Taylor expansion are known, one obtains only an approximate value for the radius of convergence. A more reliable estimate is then obtained by using the functional dependence of $R_{w}$ on $\lambda_{g}$, as described below.

In order to enhance the sensitivity to the location of the critical point and minimize the influence of other singularities, it is advantageous to use a mapping which leaves the singularity of interest close to the origin and moves all others as far away as possible~\cite{pearce}. This can be achieved by varying $\lambda_{\rm g}$ in Eq.~(\ref{lambda-g}). Thus, given 
a Taylor expansion in $\mu$, or equivalently in $\lambda$, known up to $n$-th order, we proceed by 
performing the mapping of the truncated series and obtain an expansion about $w=0$ at $n$-th order in $w$.

The radius of convergence 
as a function of the parameter $\lambda_{\rm g}$ is approximately given by\footnote{In the limit $n\to\infty$ this expression for the radius of convergence is exact.} $R^w_n = |c^{w}_n|^{(-1/n)}$.   
The approximate expression is, for a given $n$, fitted with the {\em Ansatz} $R^w(\lambda_{\rm g}) = a + w_{\rm g}(\overline{\lambda}_{\rm c};\lambda_{\rm g})$ for $R^{w}<0.95$.
The parameter $a$ accounts for corrections due the truncation at a finite $n$, while $\overline{\lambda}_{\rm c}$ is the approximate critical fugacity. We neglect the weak dependence of $a$ on $\lambda_{\rm g}$~\cite{Skokov:2010uc}.
\begin{figure}[t]
 \centerline{ \includegraphics*[width=7cm]{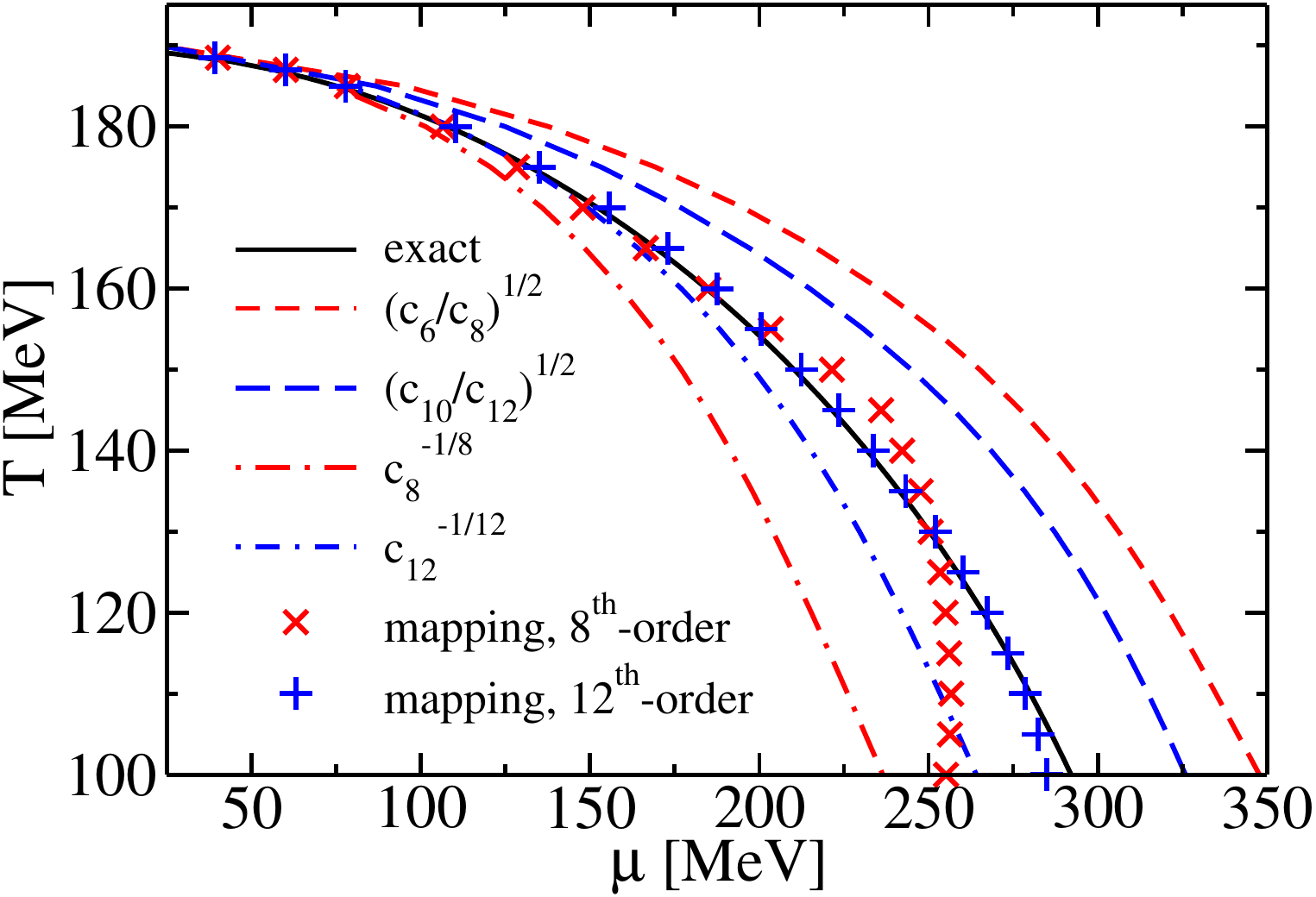}}
  \caption {The phase diagram of the QM model in the chiral limit. The  solid line represents 
	the critical line of the second-order phase transition, while the dashed 
	and dash-dotted lines are obtained using different estimates of the radius of convergence based on the original Taylor expansion in $\mu$. The symbols show the phase boundary obtained using the mapping technique discussed the text.
     }
  \label{fig:RestoredPD}
\end{figure}

The resulting fit is illustrated  in the right panel of Fig.~\ref{fig:Rconv} at a temperature $T=150$ MeV for $n=12$.
By following the procedure outlined above for temperatures in the range from 100 MeV to $T_{\rm c}$, we obtain an approximate phase diagram of the QM model shown in Fig.~\ref{fig:RestoredPD}.
For comparison, we also show the results obtained by extracting the radius of convergence directly from the original Taylor series in $\mu$, for $n=8,12$. The present approach yields a 
better estimate for the location of the critical point for all values of $n$ considered.
The fact that over a wide range of temperatures, we find approximately the same phase boundary for 
different $n$ confirms the consistency of the {\em Ansatz} for $R^{w}$.

For a physical pion mass, we expect a crossover transition at small values of the chemical potential. 
In this case the critical singularity splits into two conjugate branch points, 
located at complex values of the chemical potential symmetrically with respect to the real axis~\cite{Stephanov:2006dn,Itzykson:1983gb}, $\mu_{\rm co}$ and $\mu_{\rm co}^{\star}$, in analogy to the singularities in the $a$ plane (see Fig.~\ref{sing-a-plane}). Hence, in this case, the radius of convergence of the Taylor expansion is given by the distance of $\mu_{\rm co}$ to the origin, $|\mu_{\rm co}|$. As shown in Fig.~\ref{tmuplane_norm}, the resulting radius of convergence in the QM model decreases continuously, as the temperature is increased from the critical endpoint to the pseudo-critical temperature at $\mu=0$ and then increases again, with a minimum located close to the pseudo-critical temperature at $\mu=0$. 

In this talk, I discussed the analytic structure of thermodynamic functions in the vicinity of phase transitions and presented a scheme, employing a conformal mapping, for locating the critical point of a second-order phase transition in the thermodynamical limit. Generalizations of this approach to finite systems and to crossover transitions would be useful.

I thank Kenji Morita and Vladimir Skokov for numerous stimulating discussions. This work  was supported in part by the ExtreMe Matter Institute EMMI.

\end{document}